\documentstyle[preprint,aps,floats]{revtex}
\input epsf

\begin{document}

\font\fortssbx=cmssbx10 scaled \magstep2
\hbox to \hsize{
\hskip.5in \raise.1in\hbox{\fortssbx University of Wisconsin - Madison}
\hfill$\vtop{\hbox{\bf MADPH-97-991}
                \hbox{\bf DOE-ER40757-096}
                \hbox{\bf KEK-TH-514}
                \hbox{\bf hep-ph/9703311}
                \hbox{March 1997}}$ }

\vspace{.75in}

\begin{center}
{\large\bf Contact interactions and high-$Q^2$ events\\
 in $e^+p$ collisions at HERA}\\[10mm]
V. Barger$^1$, Kingman Cheung$^2$, K. Hagiwara$^{1,3,4}$, and D.
Zeppenfeld$^1$\\[5mm]
\it
$^1$Physics Department, University of Wisconsin, Madison, WI 53706\\
$^2$Center for Particle Physics, University of Texas, Austin, TX 78712\\
$^3$Theory Group, KEK, Tsukuba, Ibaraki 305, Japan\\
$^4$ICEPP, University of Tokyo, Hongo, Bunkyo-ku, Tokyo 113, Japan
\end{center}

\begin{abstract}
We consider $e e q q$ contact interactions as a possible origin of an
excess of events  above Standard Model
predictions in $e^+p\to e^+\rm\,jet$ at high $Q^2$ observed by the H1 and ZEUS
collaborations at HERA. The HERA data prefer chirality RL or LR contact terms
and atomic physics parity violation
measurements severely limit parity-odd contact terms.
With equal left-right and right-left chirality
interactions at an effective scale of order 3~TeV we
are able to reproduce the main features of the HERA data and still be
consistent with Drell-Yan pair production at the Tevatron and hadron
production at LEP\,2 and TRISTAN.
\end{abstract}

\thispagestyle{empty}
\newpage

The H1\cite{H1} and ZEUS\cite{zeus} experiments at HERA have observed an excess
of events in $e^+p\to e^+X$ at $Q^2>15{,}000\rm~GeV^2$ compared to Standard
Model (SM) expectations for deep inelastic scattering based on conventional
structure functions\cite{martin,lai,gluck}. In terms of the kinematic variables
$x=Q^2/2P\cdot q$ and $y=Q^2/sx$ of deep inelastic scattering, the excess
events occur at large $x$ and large $y$, $x>0.25$ and $y>0.25$. The final
states of these events are clean, with a positron and a jet back-to-back in the
transverse plane and
essentially no missing transverse 
energy. Figure~1 shows the $x,y$ distribution of the
events at $Q^2>10{,}000\rm~GeV^2$ from the two experiments. Here the $x,y$
values from the ZEUS experiment are based on the double-angle method (to
avoid uncertainties in the energy measurement); the H1 data are based on the
energy and angle of the scattered positron.  The event rate in the region
$Q^2>15{,}000\rm~GeV^2$ is about a factor of 2 above the SM prediction and
corresponds to a cross section excess of about 0.4~pb.

There are several possibilities that might explain the observed excess. First,
an upward statistical fluctuation could be responsible, although the
probability for this at $Q^2>10{,}000\rm~GeV^2$ is only 0.6\% in the H1
experiment and 0.72\% for $x>0.55$ and $y>0.25$ in the ZEUS experiment. A
second possibility is that the parton densities at large $x$ are not well
understood.
However, it is largely the valence quark distributions which are relevant here
and these are relatively well constrained in the high-$x$ region by
BCDMS\cite{bcdms}, NMC\cite{arneodo}, SLAC\cite{whitlow} and CCFR\cite{ccfr}
measurements. The ZEUS collaboration concludes that the uncertainty of this
origin is only 6.5\%. Another possibility is that large QCD effects occur near
the edge of phase space. However, a very modest $K$-factor, $K\approx
1.1$ is found\cite{MWZ} over the entire high $Q^2$ $x$-$y$ range probed by the
HERA experiments and thus there is no indication of large QCD logarithms
needing summation. Since none of the above explanations seem particularly
promising, it is natural to entertain new physics sources of the anomalous
events.

The most obvious candidate for new physics is leptoquarks (LQ) or stops with
$R$-parity violating couplings that are
produced in the $s$-channel and which decay to $e^+$ and a
quark\cite{buch,HR,kon,choudry,barg-ma,mura-yana,altarelli}. An enhancement in
the cross section would occur at a fixed $x$-value
\begin{equation}
x = M_{LQ}^2 / s \,,
\end{equation}
where $\sqrt s\simeq 300\rm~GeV$. The possibilities for $e^+p$ production of
leptoquarks or $\overlay/R$-squarks
are\cite{buch,HR,kon,choudry,barg-ma,mura-yana,altarelli}
\begin{eqnarray}
(i)&&\quad e^+u\to LQ^{+5/3}  \,,\\
(ii)&&\quad e^+d\to \tilde t^{+2/3}  \,,\\
(iii)&&\quad e^+\bar u\to \bar D^{+1/3}  \,.\\
(iv)&&\quad e^+\bar d\to LQ^{+4/3}
\end{eqnarray}
Here $\tilde t$ is the stop particle of supersymmetry and 
$\bar D$ is the leptoquark state in the 27-dimensional representation of
$E_6$. The cross-section for leptoquark production is
\begin{equation}
\sigma(e^+p\to LQX) = \left(4\pi^2\over s\right)f_{q/p}(x, M_{LQ}^2)
{\Gamma(LQ\to e^+q)\over
M_{LQ}} (2J+1)\,,
\end{equation}
where $f_{q/p}$ is the quark distribution in the proton and $J$ is the spin of
the leptoquark. From the excess in the $e^+p\to e^+X$ cross section,
$\Delta\sigma\approx 0.4~$pb, we can calculate the leptoquark decay width and
estimate its coupling strength. For the above four
scenarios, using MRSA structure functions\cite{martin} and $M_{LQ} = 200$~GeV,
we obtain for $J=0$
\begin{equation}
\arraycolsep=1em
\begin{array}{rcc}
 & B\,\Gamma\rm\ (MeV)& \sqrt B\,\lambda_{LQ}/e\\
(i)& 1.4 & 0.06 \\
(ii)& 5.7 & 0.12\\
(iii)& 350 & 1.0\\
(iv)& 130 & 0.6
\end{array}
\end{equation}
where $e=\sqrt{4\pi\alpha}$, $\Gamma$ is the total leptoquark width and $B$ is
the branching fraction to the $eq$ final
state. Since the leptoquark width would be narrow, all of the excess events
should fall in a single $x$ bin, within resolution, in this $s$-channel
resonance scenario.
Although the H1 data above are somewhat suggestive of a resonant enhancement at
$M\approx 200$~GeV (see the vertical dotted line in Fig.~1), this
interpretation is not indicated by the combined  H1
and ZEUS data. Of course, two or more leptoquark resonances
could exist with similar but non-degenerate masses which could lead to a
broader $x$-distribution. Since standard leptoquark and $R$-parity
breaking\cite{kon,choudry} stop scenarios have been discussed
extensively elsewhere, we
do not pursue these interesting possibilities further here.

A final interesting potential explanation of the HERA high-$Q^2$ phenomena is
new physics in the $t$-channel. Any particles beyond the photon and $Z$-boson
that can be exchanged in the $t$-channel of $e^+p\to e^+X$ must have TeV mass
scale to be consistent with present measurements of $e^+e^-\to q\bar q$ and
$p\bar p\to e^+e^- X$. Thus models of this type can be generically
parameterized by a contact interaction\cite{altarelli,ELP,cashmore,chiap}.

The conventional effective Lagrangian of an $e e q q$ contact
interaction has the form\cite{ELP,cashmore,chiap}
\begin{eqnarray}
L_{NC} &=& \sum_q \Bigl[ \eta_{LL}\left(\overline{e_L} \gamma_\mu e_L\right)
\left(\overline{q_L} \gamma^\mu q_L \right) + \eta_{RR} \left(\overline{e_R}
\gamma_\mu e_R\right) \left( \overline{q_R}\gamma^\mu q_R\right) \nonumber\\
&& \qquad {}+ \eta_{LR} \left(\overline{e_L} \gamma_\mu e_L\right)
\left(\overline{q_R}
\gamma^\mu q_R\right) + \eta_{RL} \left(\overline{e_R} \gamma_\mu e_R\right)
\left(\overline{q_L} \gamma^\mu q_L \right) \Bigr] \,, \label{effL}
\end{eqnarray}
where the coefficients have dimension (TeV)$^{-2}$ and are conventionally
expressed as $\eta_{\alpha\beta}^{eq} = \epsilon 4\pi /\Lambda_{eq}^2$, where
$\eta_{\alpha\beta}^{eu}$ and $\eta_{\alpha\beta}^{ed}$ are independent
parameters. Here $\epsilon= \pm 1$ allows
for either constructive or destructive interference with the SM $\gamma$ and
$Z$ exchange amplitudes and $\Lambda_{eq}$ is the effective mass scale of the
contact interaction.
In the effective interaction (\ref{effL}) we do not
include lepton or quark chirality violating terms  such as $\overline{e_L}
 e_R \overline{q_L} q_R$.
This contact interaction can arise from particle exchanges in $s$, $t$, or $u$
channels for which the mass-squared of the exchanged particle is much larger
than the corresponding Mandelstam invariant.
A contact interaction may also be a manifestation of
fermion compositeness\cite{ELP}.
Contact interactions which arise from the exchange of a single heavy particle
will factorize into two vertex factors.
An example of a $t$-channel exchange is a heavy vector boson $Z'$, which gives
\begin{equation}
\eta_{\alpha\beta} = -g_\alpha^e g_\beta^q / M_{Z'}^2 \,, \label{Z'}
\end{equation}
where $g_\alpha^e$ and $g_\beta^q$ are the $Z'$ couplings to $e_\alpha$ and
$q_\beta$, respectively.  In this case  factorization implies
additional contact interactions for $ ee ee$ and $ qq qq$. If
the interactions satisfy lepton universality there will be further contact
amplitudes for $ ee\mu\mu$, $ ee\tau\tau$, etc. and, since  couplings to
neutrinos are expected in general, $\nu\nu
ee$ and $\nu\nu qq$ contact interactions would also exist. There may
also be corresponding contact interactions for charged currents such as $\nu
e d u$ corresponding to heavy $W$-bosons. Thus a rich phenomenology
could be associated with contact amplitudes representing new physics beyond the
energy reach of present colliders.

  The reduced amplitudes for $eq\to eq$, $\bar qq\to e^+e^-$, and
$e^+e^-\to \bar qq$  subprocesses, including Standard Model plus contact
interactions, are given by
\begin{equation}
M_{\alpha\beta}^{eq} = {e^2 Q_e Q_q\over\hat t} + {g_Z^2 (T_{e\alpha}^3 - 
s^2_{\rm w}Q_e) (T_{q\beta}^3 - s^2_{\rm w} Q_q) \over \hat t - m_Z^2} +
\eta_{\alpha\beta}^{eq} \,, \label{amps}
\end{equation}
where $Q_f$ and $T_{f\alpha}^3$ are charges and weak isospins, respectively, of
the external fermions $f_\alpha$, $g_Z = e/(\sin\theta_{\rm w} \, 
\cos\theta_{\rm w})$,
$s_{\rm w}=\sin\theta_{\rm w}$, and the
Mandelstam invariant $\hat t$ is given by
\begin{eqnarray}
&& \hat t = (E_{\rm c.m.})^2 \hspace{.725in} {\rm for}\ e^+e^- \,,\nonumber\\
&& \hat t = -Q^2 = -sxy \qquad {\rm for}\ e^\pm p \,,\\
&& \hat t  = sx_1x_2 \hspace{.85in} {\rm for}\ p\bar p \,.\nonumber
\end{eqnarray}
Thus the presence of a contact interaction has interconnected implications for
$ep\to eX$, $p\bar p\to e^+e^-X$, $e^+e^-\to \rm hadrons$ and atomic physics
parity violation experiments.
The cross sections are related to the amplitudes (\ref{amps}) and parton
densities $u$ and $d$ as follows:
\begin{eqnarray}
{d\sigma(e^+p)\over dx\,dy} &=& {sx\over16\pi} \left\{ u(x,Q^2) \left[
\left| M_{LR}^{eu} \right|^2 + \left| M_{RL}^{eu} \right|^2 + (1-y)^2 \left(
\left| M_{LL}^{eu} \right|^2 + \left| M_{RR}^{eu} \right|^2 \right) \right]
\right. \nonumber\\
&&\qquad\left. {}+ d(x,Q^2) \left[ \left| M_{LR}^{ed} \right|^2 + \left|
M_{RL}^{ed} \right|^2 + (1-y)^2 \left( \left| M_{LL}^{ed} \right|^2 + \left|
M_{RR}^{ed} \right|^2 \right) \right] \right\}  \label{bar-ep}\\
{d\sigma(e^-p)\over dx\,dy} &=& {sx\over16\pi} \left\{ u(x,Q^2) \left[ \left|
M_{LL}^{eu} \right|^2 + \left| M_{RR}^{eu} \right|^2 + (1-y)^2 \left( \left|
M_{LR}^{eu} \right|^2 + \left| M_{RL}^{eu} \right|^2 \right) \right] \right.
\nonumber\\
&&\qquad\left. {}+ d(x,Q^2) \left[ \left| M_{LL}^{ed} \right|^2 + \left|
M_{RR}^{ed} \right|^2 + (1-y)^2 \left( \left| M_{LR}^{ed} \right|^2 + \left|
M_{RL}^{ed} \right|^2 \right) \right] \right\}
\end{eqnarray}
\begin{eqnarray}
{d\sigma(p\bar p\to \ell\bar\ell X)\over dx_1 dx_2 d\cos\hat\theta} &=&  {\hat
s\over 384\pi} \left\{ (1+\cos\hat\theta)^2 \left[ u(x_1,\hat s) u(x_2,\hat s)
\left( \left| M_{LL}^{eu} \right|^2 + \left| M_{RR}^{eu} \right|^2 \right)
\right. \right.
\nonumber\\
&&\hspace{1.35in} \left. {}  + d(x_1,\hat s) d(x_2,\hat s) \left( \left|
M_{LL}^{ed} \right|^2 + \left| M_{RR}^{ed} \right|^2 \right) \right]
\nonumber\\
&&\qquad {}+ (1-\cos\hat\theta)^2 \left[ u(x_1,\hat s) u(x_2,\hat s) \left(
\left| M_{LR}^{eu} \right|^2 + \left| M_{RL}^{eu} \right|^2 \right) \right.
\nonumber\\
&&\hspace{1.35in} \left. \left. {} + d(x_1,\hat s) d(x_2,\hat s) \left( \left|
M_{LR}^{ed} \right|^2 + \left| M_{RL}^{ed} \right|^2 \right) \right] \right\}
\end{eqnarray}
\begin{equation}
{d\sigma(e^+ e^-\to q\bar q)\over d\cos\theta} = {3s\over 128\pi} \sum_q
\left\{ (1+\cos\theta)^2 \left( \left| M_{LL}^{eq} \right|^2 + \left|
M_{RR}^{eq} \right|^2 \right) + (1-\cos\theta)^2 \left( \left| M_{LR}^{eq}
\right|^2 + \left| M_{RL}^{eq} \right|^2 \right) \right\} \,.
\end{equation}

A parity non-conserving contact interaction would modify the SM prediction for
the atomic physics parity violating parameter $Q_W$, given by
\begin{equation}
Q_W = Q_W^{\rm SM} - {1\over \sqrt{2}G_F} \left[ (N+2Z)\Delta\eta^{eu} +
(2N+Z)\Delta\eta^{ed} \right] \,,
\end{equation}
where $N$ is the number of neutrons, $Z$ is the number of protons, and
\begin{equation}
\Delta\eta^{eq} = \eta_{RR}^{eq}-\eta_{LL}^{eq}+\eta_{RL}^{eq}-\eta_{LR}^{eq}
\,.
\end{equation}
The ${}^{133}_{\phantom055}$Cs measurements find $Q_W =  - 71.04 \pm 1.81$
while the SM prediction for $m_t = 175$~GeV and $m_H = 100$~GeV is $Q_W^{\rm
SM} = -73.04$. The difference $\Delta Q_W = 2.0 \pm 1.8$ places a severe
constraint on allowable contact interactions,
\begin{equation}
\Delta Q_W = (11.4\; {\rm TeV}^2 )
\left( \eta_{LL}^{eu} + \eta_{LR}^{eu} - \eta_{RL}^{eu} -
\eta_{RR}^{eu} \right) + (12.8\; {\rm TeV}^2)
 \left(\eta_{LL}^{ed} + \eta_{LR}^{ed} -
\eta_{RL}^{ed} - \eta_{RR}^{ed}\right) \,.
\end{equation}
Parity conserving contact interactions such as $\eta_{LR}^{eq} =
\eta_{RL}^{eq}$ and $\eta_{LL}^{eq} =
\eta_{RR}^{eq}$ give $\Delta Q_W = 0$.

In order to develop some feeling for the interference of the SM and contact
amplitudes, we give numerical values for the SM chirality amplitudes in
Table~\ref{table:SM-amps}, at relevant values of $\hat t$, namely $\hat t =
-20{,}000\rm~GeV^2$ for $eq\to eq$ and $\hat t = (175\rm~GeV)^2$ for $\bar
qq\to e^+e^-$ or $e^+e^-\to q\bar q$. The SM amplitude changes
sign under crossing, because $1/\hat t$ changes sign
in the dominant photon amplitude, 
but the contact terms have the same sign in
both direct and crossed channel amplitudes. This has the consequence that
constructive interference in an $eq\to eq$ amplitude corresponds with
destructive interference in the $q\bar q\to  e^+e^-$ amplitude. Moreover,
$ep$ cross sections may be more or less sensitive to the presence of the
contact amplitude than $e^+e^-$ or $\bar pp$, depending on the relative size of
the SM contributions and the sign of the contact contribution.

\begin{table}
\caption{Chirality amplitudes for $e^-q\to e^-q$ at $\hat t =
-20{,}000\rm~GeV^2$ and for $\bar qq\to e^+e^-$ or $e^+e^-\to \bar qq$ at $\hat
t = (175\rm~GeV)^2$, where $q=u,d$. The amplitude units are
(TeV)$^{-2}$.\label{table:SM-amps}}
\arraycolsep=1em
\[
\begin{array}{|l|c|c|c|c|}
\hline
& eu\to eu & ed\to ed & \bar uu\to e^+e^- & \bar dd\to  e^+e^- \\
\hline
LL & 5.1 + \eta_{LL}^{eu} & -3.8 + \eta_{LL}^{ed} & -4.4 +
\eta_{LL}^{eu} & 3.9 + \eta_{LL}^{ed}\\
RR & 3.9 + \eta_{RR}^{eu} & -2.0 + \eta_{RR}^{ed} & -3.0 +
\eta_{RR}^{eu} & 1.5 + \eta_{RR}^{ed} \\
LR & 2.4 + \eta_{LR}^{eu} & -1.2 + \eta_{LR} & -1.1 +
\eta_{LR}^{eu} & 0.6 + \eta_{LR}^{ed}\\
RL & 1.7 + \eta_{RL}^{eu} & 0.3 + \eta_{RL}^{ed} & -0.2 +
\eta_{RL}^{eu} & -1.3 + \eta_{RL}^{ed}
\\ \hline
\end{array}
\]
\end{table}

{}From (\ref{bar-ep}) we see that a
high-$y$ anomaly in $e^+p$ requires an $\eta_{LR}^{eq}$ or $\eta_{RL}^{eq}$
amplitude that interferes constructively with the SM contribution.
$\eta_{LL}^{eq}$ or $\eta_{RR}^{eq}$ amplitudes are suppressed by the $(1-y)^2$
factor in $e^+p$ collisions but would be enhanced compared to
$\eta_{LR}$ or $\eta_{RL}$ terms in $e^-p$ scattering.
Because the $d$-parton density is severely suppressed in the large $x$ region
relative to the $u$-parton density\cite{martin,lai,gluck},
the $eu$ contact interaction is needed to achieve an $e^+p$ cross section
enhancement. As an illustrations we consider the following scenarios
for the contact contributions:
\begin{eqnarray}
&&\eta_{RL}^{eu} =\eta_{LR}^{eu} =  1.4\rm\ TeV^{-2}
\,,\label{contrib-a} \\
&&\eta_{RL}^{eu} = -\eta_{LR}^{eu} =  2.6 \rm\ TeV^{-2} \,.
\label{contrib-b}
\end{eqnarray}
These correspond to effective scales $\Lambda_{eu}=3$~TeV and
$\Lambda_{eu}=2.2$~TeV,
respectively.

The effect on the HERA cross sections is demonstrated in Figs.~2 and 3.
In Fig.~2 we show the $e^+p$ DIS cross section, $\sigma(Q^2 > Q^2_{\rm min})$,
as a function of the minimal $Q^2$ value; also included are the HERA
results from Refs.~\cite{H1,zeus}, corrected for detection efficiencies
of 80\% and 81.5\%, respectively, and divided by an average QCD K-factor
of 1.1 since we are showing leading order cross sections throughout.
Both choices give better representations of the data than the SM at $Q^2 >
15{,}000\rm~GeV^2$. The second
scenario was chosen to introduce a destructive interference with SM amplitudes
at $Q^2_{\rm min}\approx 10,000$~GeV$^2$.
The corresponding SM and contact interaction contributions to $d\sigma/dx$
and $d\sigma/dy$ for $e^+p\to e^+X$ are shown in Figure~3. The features of
the data in Fig.~1 are qualitatively reproduced by
these choices of contact terms.

Of the two scenarios, the first
case is parity conserving, with a $VV-AA$ interaction in
(\ref{contrib-a}), where $V$ denotes vector and $A$ axial vector, and thus
satisfies the $Q_W$-constraint. The choice in (\ref{contrib-b}) improves
the description of the HERA  $Q^2$ distribution over (\ref{contrib-a})
but it gives a contribution $\Delta Q_W(\rm contact)= -59.3$ that is in severe
conflict with the measured value. This $Q_W$-discrepancy can
be rectified by introducing additional, cancelling $\eta^{eu}_{LL}$ or
$\eta^{eu}_{RR}$ contributions or contact terms involving $d$-quarks. Such a
change would affect the HERA $e^+p$ cross sections very little, due to the
$(1-y)^2$ terms in Eq.~(\ref{bar-ep}) and the suppressed down-quark density.

Figure 4 compares the SM prediction for the Drell-Yan cross section at the
Tevatron with the calculated cross section for the contact amplitudes of
(\ref{contrib-a}) and (\ref{contrib-b}). We see that the contact amplitudes
in (\ref{contrib-a}) are consistent with the preliminary CDF\cite{bodek} data
but the choice in (\ref{contrib-b}) is ruled out.
Any attempt to resolve the $Q_W$-discrepancy of scenario (\ref{contrib-b})
by additional contact terms will further worsen the disagreement with the
CDF data. The changes in
$\sigma(e^+e^-\to\rm hadrons)$ caused by (\ref{contrib-a}) and
(\ref{contrib-b}) are sufficiently small as not to conflict with
current LEP\,2 measurements~\cite{altarelli,opal}.

In summary we have demonstrated the following:
\begin{enumerate}
\item
The presence of an $eeuu$ contact term
could satisfactorily explain the observed excess of events at high $Q^2$
in the HERA H1 and ZEUS experiments.
\item
A new physics scale $\Lambda \alt 3$~TeV
in LR and/or RL amplitudes is required.
\item
The CDF Drell-Yan data allow a
contact term of this order, as long as it occurs in only one or two of the
eight chirality amplitudes for $u \bar u \to e^+e^-$ and
$d\bar d \to e^+e^-$.
\item
A destructive
interference would cause a rapid onset of the new physics contributions with
increasing $Q^2$ which would improve the agreement with the shape of the $Q^2$
distribution at HERA. However, because of the tightness of the Tevatron bounds
on $eeqq$ contact terms there is little room to tolerate such destructive
interference.
\item
Atomic physics parity violation measurements severly constrain parity-odd
contact terms.
\item
Deeply inelastic $e^-p$ scattering is not sensitive to the LR and RL terms
needed to explain the $e^+p$ data but would be complementary in probing
LL and RR contact terms. However, the presence of additional chirality terms
would exacerbate
the situation with the Drell-Yan cross section at high lepton-pair mass.
\end{enumerate}

\section*{Acknowledgments}
K.H. would like to thank G.C.~Cho, S.~Matsumoto and Y.~Umeda for discussions.
This research was supported in part by the U.S.~Department of Energy under
Grant Nos. DE-FG03-93ER40757 and 
DE-FG02-95ER40896 and in part by the University of Wisconsin Research
Committee with funds granted by the Wisconsin Alumni Research Foundation.

\newpage
\section*{Figure Captions}

\begin{enumerate}

\item[Fig.~1:]
Event distributions at $Q^2>10{,}000\rm~GeV^2$ in $e^+p\to e^+X$ from the H1
(solid points) and ZEUS (open points) experiments.

\item[Fig.~2:]
Integrated cross sections versus a minimum $Q^2$ for $e^+p\to e^+X$
for the SM (solid curve) and the contact interactions of
(\protect\ref{contrib-a}) (dashed curve) and (\protect\ref{contrib-b})
(dotted curve). The data points are combined H1 and ZEUS measurements.

\item[Fig.~3:]
Predicted $x$ and $y$ distributions for $Q^2>15{,}000\rm~GeV^2$ and $y>0.2$
for
the SM (solid curves) and with two choices of the contact interactions of
(\protect\ref{contrib-a}) (dashed curves) and (\protect\ref{contrib-b})
(dotted curves).

\item[Fig.~4:]
The Drell-Yan cross section $\bar pp\to \mu^+\mu^-X + e^+e^-X$ at the Tevatron
($\protect\sqrt s = 1.8$~TeV) for the SM with the contact interactions of
(\protect\ref{contrib-a}) (dashed curve) and (\protect\ref{contrib-b})
(dotted curve). Preliminary CDF data from
Ref.~\protect\cite{bodek} are compared. A constant $K$-factor is determined
from the data in the region of the $Z$-resonance.
\end{enumerate}

\newpage
\begin{figure}[t]
\centering
\leavevmode
\epsfysize=460pt
\epsfbox{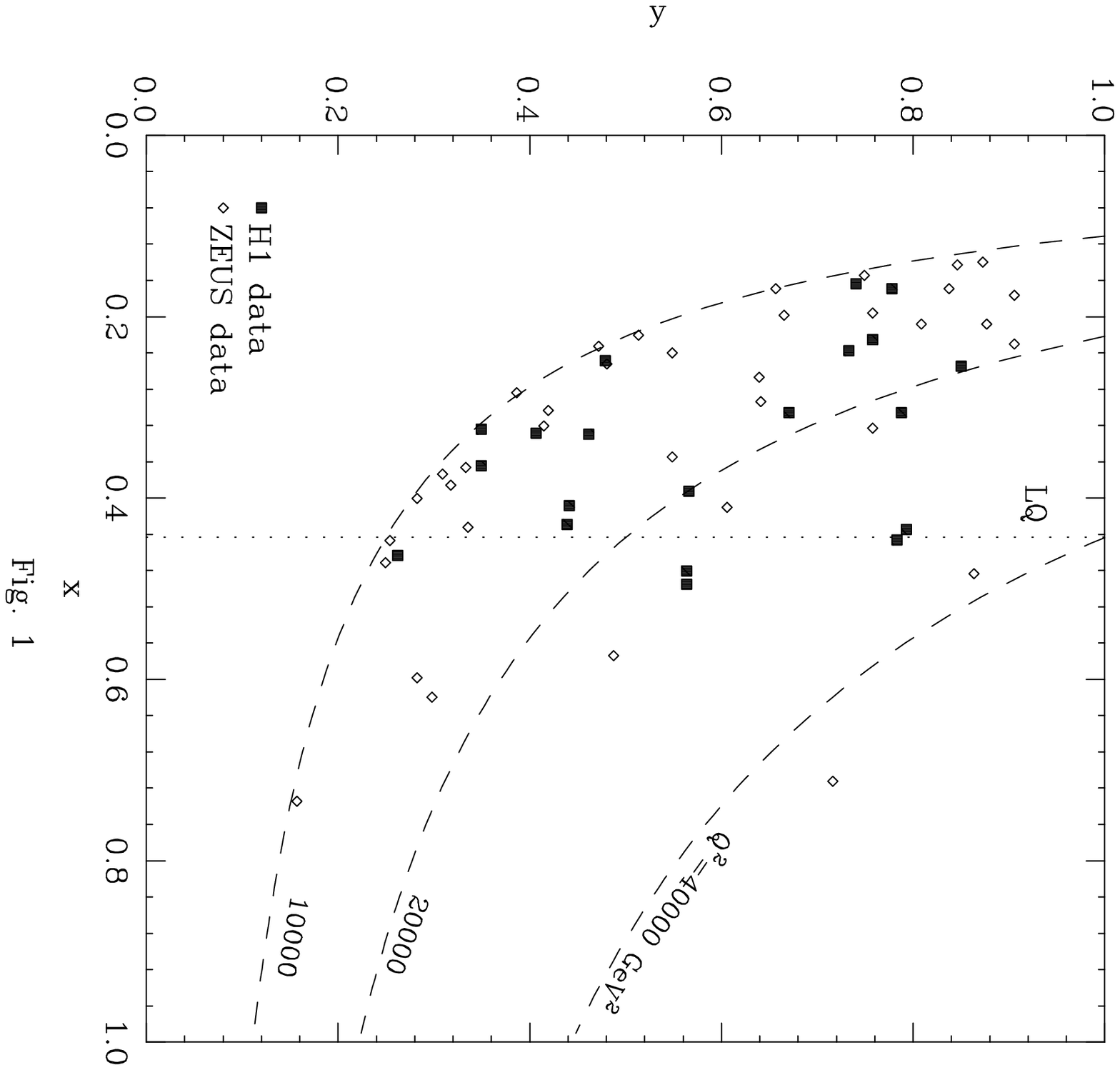}
\end{figure}

\begin{figure}[t]
\centering
\leavevmode
\epsfysize=580pt
\epsfbox{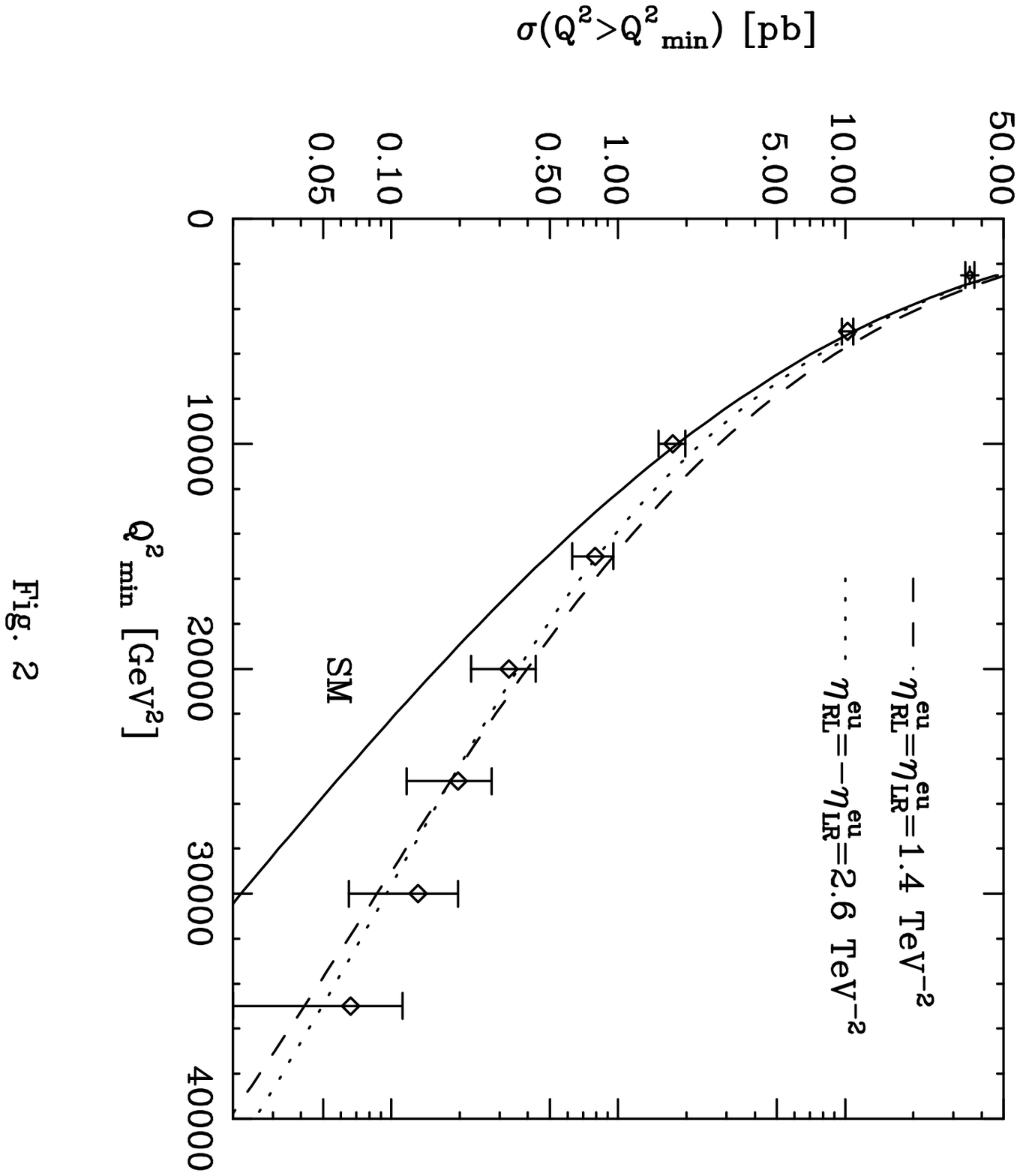}
\end{figure}

\begin{figure}[t]
\centering
\leavevmode
\epsfysize=580pt
\epsfbox{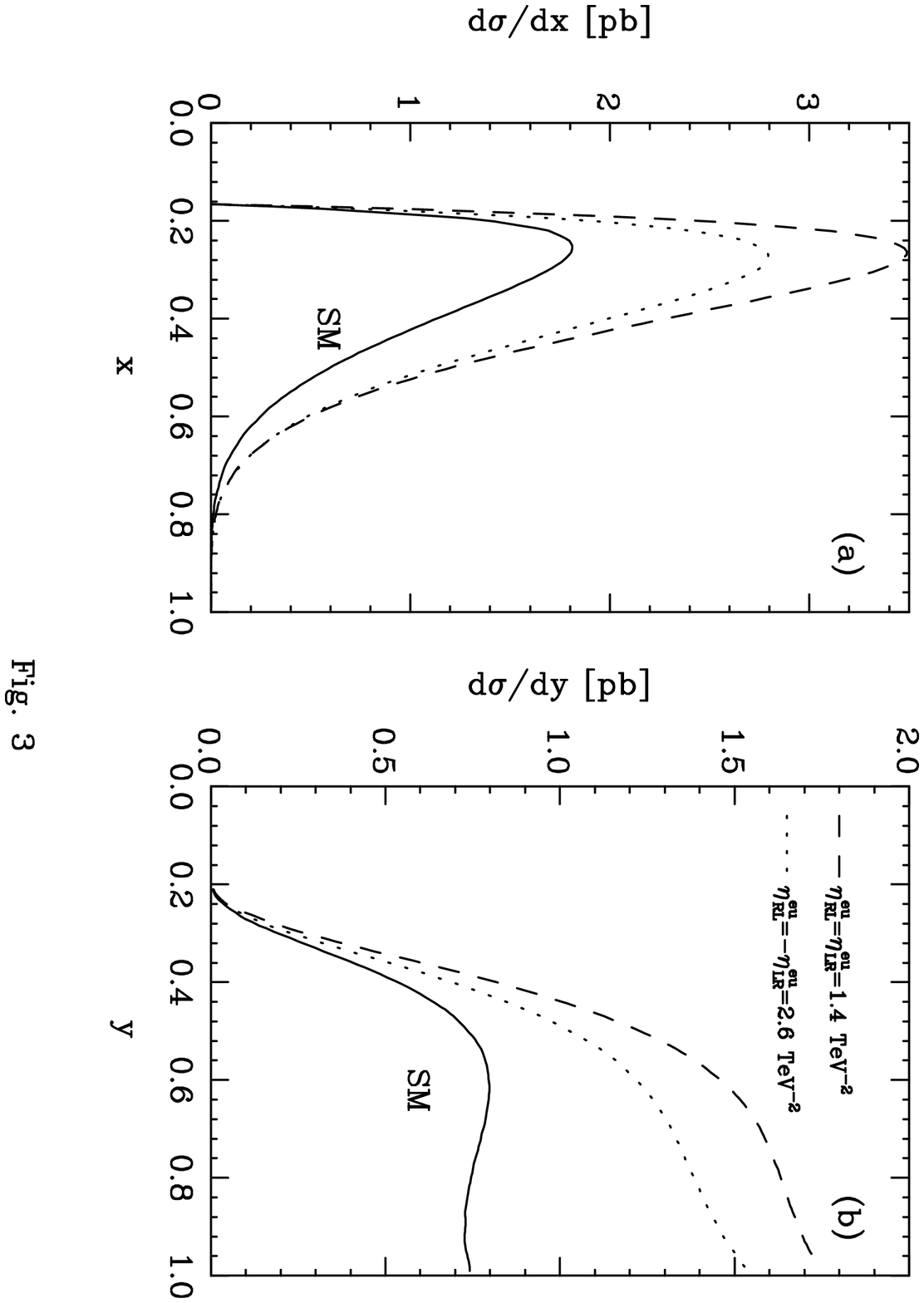}
\end{figure}

\begin{figure}[t]
\centering
\leavevmode
\epsfysize=580pt
\epsfbox{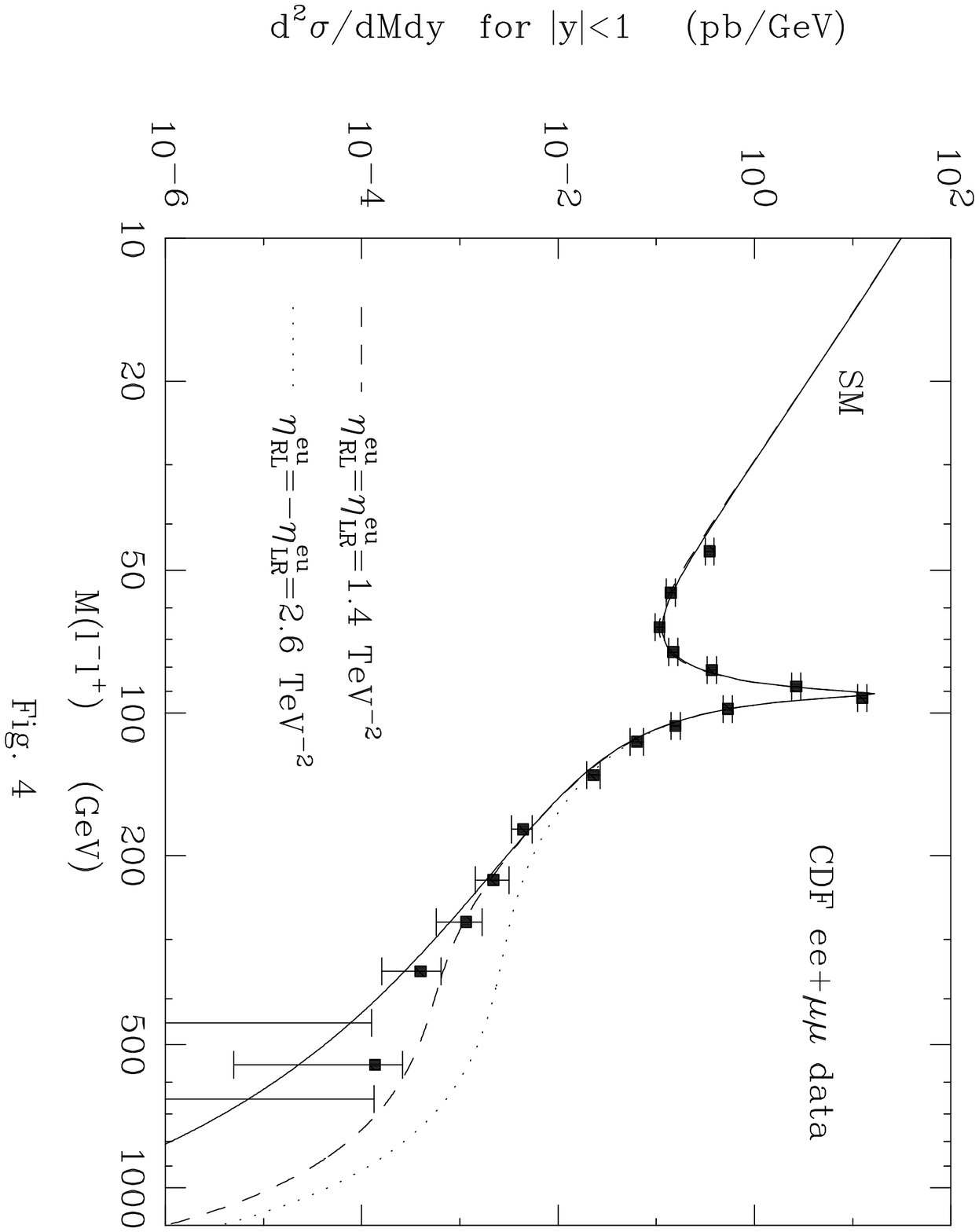}
\end{figure}


\begin{references}

\bibitem{H1} The H1 Collaboration, C. Adloff et al., report DESY 97-24
[hep-ex/9702012] (1997).

\bibitem{zeus} The ZEUS Collaboration, J. Breitweg et al., report DESY 97-025
[hep-ex/9702015] (1997).

\bibitem{martin} A.D. Martin, W.J. Stirling, and R.G. Roberts, Phys.\ Rev.\
{\bf D59}, 6734 (1994); Int.~J. Mod.\ Phys.\ {\bf A10}, 2885 (1995); Phys.\
Lett.\ {\bf B387}, 419 (1996).

\bibitem{lai} H.L. Lai et al., Phys.\ Rev.\ {\bf D51}, 4763 (1995); {\bf D55},
1280 (1997).

\bibitem{gluck} M. Gl\"uck, E.~Reya, and A.~Vogt, Z.~Phys.\ {\bf C67}, 433
(1995).

\bibitem{bcdms} BCDMS Collaboration: A.C.~Benvenuti et al., Phys.\ Lett.\ {\bf
B223}, 485 (1989); {\bf B237}, 592 (1990).

\bibitem{arneodo} M. Arneodo et al., Phys.\ Lett.\ {\bf B309}; hep-ph/9610231
(Oct.~1996), submitted to Nucl.\ Phys.

\bibitem{whitlow} L.W. Witlow et al., Phys.\ Lett.\ {\bf B282}, 475 (1992).

\bibitem{ccfr} CCFR Collaboration: P.Z.~Quintas et al., Phys.\ Rev.\ Lett.\
{\bf 77}, 438 (1996).

\bibitem{MWZ} E.~Mirkes, S.~Willforth, and D.~Zeppenfeld (unpublished).

\bibitem{buch} W.~B\"uchmuller, R.~Rueckl, and D.~Wyler, Phys.\ Rev.\
Lett.\ {\bf 191}, 442 (1987); M.~Leurer, Phys.\ Rev.\ {\bf D49}, 333 (1994);

\bibitem{HR}
J. Gunion and E. Ma, Phys.\ Lett.\ {\bf B195}, 257 (1987); V.~Barger,
K.~Hagiwara, T.~Han, and D.~Zeppenfeld, Phys.\ Lett. {\bf B220}, 464 (1989);
J.~Hewett and T.~Rizzo, Phys.\ Rept.\ {\bf 183}, 193 (1989).

\bibitem{kon}
T. Kon and T. Kobayashi, Phys.\ Lett.~B {\bf 270}, 81 (1991); T.~Kon,
T.~Kobayashi, and S.~Kitamura, Phys.\ Lett.~B {\bf 333}, 263 (1994).

\bibitem{choudry} D. Choudhury and S. Raychaudhuri, report CERN-TH-97-26
[hep-ph/9702392]; J.~Bl\"umlein, report DESY~97-032 [hep-ph/9703287];
J.~Kalinowski, R.~R\"uckl, H.~Spiesberger, and P.~Zerwas, [hep-ph/9703288];
H.~Dreiner and P.~Morawitz, hep-ph/9703279.


\bibitem{barg-ma} V. Barger and E. Ma, Phys.\ Rev.\ {\bf D51}, 1332 (1995).

\bibitem{mura-yana} H. Murayama and T. Yanagida, Mod.\ Phys.\ Lett.\ {\bf A7},
147 (1992).

\bibitem{altarelli} G. Altarelli, J. Ellis, G.F. Guidice, S. Lola, and M.L.
Mangano, report CERN-TH/97-40 [hep-ph/9703276].

\bibitem{ELP} E.~Eichten, K.~Lane, and M.~Peskin, Phys.\ Rev.\ Lett.\ {\bf 50},
811 (1982).

\bibitem{cashmore} R.J. Cashmore et al., Phys.\ Rev.\ {\bf 122}, 275 (1985);
R.~R\"uckl, Phys.\ Lett.\ {\bf B129}, 363 (1983); Nucl.\ Phys.\ {\bf B234}, 91
(1984).

\bibitem{chiap} P. Chiapetta and J.-M. Virey, Phys.\ Lett.\ {\bf B389}, 89
(1996).

\bibitem{bodek}
A.~Bodek for the CDF Collaboration, preprint Fermilab-Conf-96/341-E (1996).

\bibitem{opal}
OPAL Collaboration, G.~Alexander et al., CERN-PPE/96-156.

\end{references}
\end{document}